\documentclass[floats,aps,showpacs,twocolumn]{revtex4}

\usepackage[dvips]{graphicx}
\usepackage{amsfonts}

\newcommand{\heff}{h_{\text{eff}}}
\newcommand{\Areg}{A_{\text{reg}}}
\newcommand{\Wch}{W_{\text{ch}}}
\newcommand{\freg}{f_{\text{reg}}}
\newcommand{\tH}{\tau_{\text{\tiny{H}}}}
\newcommand{\Dch}{\Delta_{\text{ch}}}
\newcommand{\mmax}{m_{\text{max}}}
\newcommand{\mstar}{m^{*}}

\newcommand{\PSImagx}[2]{\includegraphics[width=#2]{#1}}

\newcommand{\Z}{\mathbb{Z}}

\begin{document}

\title{Flooding of Chaotic Eigenstates into Regular Phase Space Islands}

\author{Arnd B\"acker, Roland Ketzmerick, and Alejandro G. Monastra}

\affiliation{Institut f\"ur Theoretische Physik, Technische
Universit\"at Dresden, 01062 Dresden, Germany}

\date{\today}

\begin{abstract}
We introduce a criterion for the existence of regular states in
systems with a mixed phase space. If this condition is not fulfilled
chaotic eigenstates substantially extend into a regular island. Wave
packets started in the chaotic sea progressively flood the island. The
extent of flooding by eigenstates and wave packets increases
logarithmically with the size of the chaotic sea and the time,
respectively. This new effect is observed for the example of
island chains with just 10 islands.
\end{abstract}
\pacs{05.45.Mt, 03.65.Sq}

\maketitle

One of the cornerstones in the understanding of the structure of
eigenstates in quantum systems is the semiclassical eigenfunction
hypothesis\ \cite{Scefct}: in the semiclassical limit the eigenstates
concentrate on those regions in phase space which a typical orbit
explores in the long-time limit. For integrable systems these are the
invariant tori. For ergodic dynamics the eigenstates become
equidistributed on the energy shell \cite{Qerg}. Typical systems have
a mixed phase space, where regular islands and chaotic regions
coexist. In this case the semiclassical eigenfunction hypothesis
implies that the eigenstates can be classified as being either regular
or chaotic according to the phase-space region on which they
concentrate. Note, that this may fail for an infinite phase space
\cite{HufKetOttSch2002}.

\begin{figure}[b]
  \begin{center}
    \PSImagx{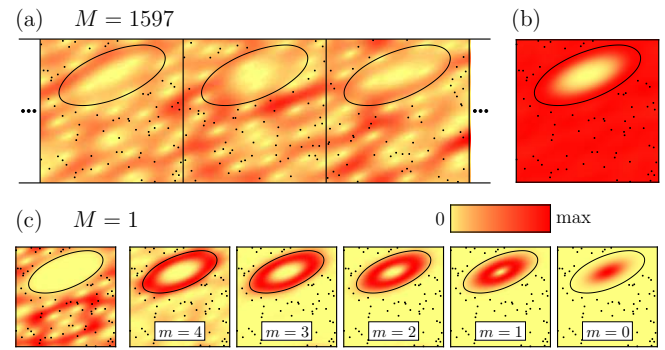}{8.6cm}
    \caption{(a) Husimi representation of a chaotic state flooding the
    regular islands. Shown are three phase-space cells out of
    $M=1597$. Full curves are classical tori close to the border of
    the regular islands and black dots are iterates of a chaotic
    orbit. This eigenstate extends well into the islands, while having
    no weight in their central region. (b) Average of the eigenstate
    over all $M$ cells. (c) For $M=1$ eigenstates concentrate
    either on the chaotic component (left) or over the $m$-th
    quantized regular torus.  For all plots $\heff \approx 1/30$ is
    used.
    \label{fig:Husimis} } \end{center}
\end{figure}

In this paper we study mixed systems with a compact phase space, but
away from the semiclassical limit. Here the properties of eigenstates
depend on the size of phase-space structures compared to Planck's
constant $h$. In the case of 2D maps this can be very simply stated
\cite{BerBalTabVor79}: a regular state with quantum number $m=0,1,...$
will concentrate on a torus enclosing an area $(m+1/2) h$, as can be
seen in Fig.~\ref{fig:Husimis}(c).

We will show that this WKB-type quantization rule is not a sufficient
condition. We find a second criterion for the existence of a regular
state on the $m$-th quantized torus,
\begin{equation}  \label{newcondition}
  \gamma_m < \frac{1}{\tH} . 
\end{equation}
Here $\tH = h / \Dch$ is the Heisenberg time of the chaotic sea with
mean level spacing $\Dch$ and $\gamma_m$ is the decay rate of the
regular state $m$ if the chaotic sea were infinite. Quantized tori
violating this condition will not support regular states. Instead,
chaotic states will {\it flood} these regions, see
Fig.~\ref{fig:Husimis}(a). In terms of dynamics we find that wave
packets started in the chaotic sea progressively flood the island as
time evolves. Partial and even complete flooding is possible,
depending on system properties. These findings are relevant for
islands surrounded by a large chaotic sea.

We numerically demonstrate the flooding and the disappearance of
regular states for the important case of island chains. In typical
Hamiltonian systems they appear around any regular island. On larger
scales they are relevant for Hamiltonian ratchets
\cite{SchOttKetDit2001}, the kicked rotor with accelerator modes
\cite{Izr90}, and the experimentally
\cite{SteOskRai2001,Hen2001,expAccel} and theoretically
\cite{theoryAccel} studied kicked atom systems. The flooding of
regular islands by chaotic states is a new quantum signature of a
classically mixed phase space. This phenomenon shows that not only
local phase-space structures, but also global properties of the phase
space determine the characteristics of quantum states.

Before we explain the origin of Eq.~(\ref{newcondition}), we
numerically study its consequences. We choose a system, where we can
change $\tH$ by increasing the system size without affecting the rates
$\gamma_m$. A one dimensional kicked system
\begin{equation}
  H(p,x,t) =  T(p) + V(x) \sum_n \delta(t-n) , \label{hamiltonian}
\end{equation}
has a stroboscopic time-evolution given by the mapping, $x_{t+1}= x_t
+ T'(p_t), \ p_{t+1} = p_t-V'(x_{t+1})$. The phase space is compact with
periodic boundary conditions for $x_t\in [0,M]$ and
$p_t\in[-1/2,1/2]$. Choosing the functions $V'(x)$ and $T'(p)$ 
appropriately \cite{fctchoice} we get a chain of $M$ islands, one per
unit cell (see Fig.~\ref{fig:Husimis}). The islands cover a relative area
$\Areg=0.215$ and have fine structure close to their boundary that is
negligible for the quantum properties studied here. Points inside an
island are mapped one unit cell to the right, i.e.\ the island chain
is transporting.

The eigenstates $|\psi\rangle$ of the quantum system are determined by
the eigenvalue equation, $\hat{U}|\psi\rangle = e^{i \varphi}
|\psi\rangle$, where $\hat U$ is the unitary time-evolution operator
over one time period, 
$\hat{U}= \exp[-2\pi i T(\hat{p})/\heff]\,\exp[-2\pi i V(\hat{x})/\heff]$. 
The spatial periodicity after $M$
cells requires an effective Planck's constant $\heff=M/N$, with
incommensurate integers $M$ and $N$. We choose for $M/N$ the rational
approximants of $\heff=1/(d+\sigma)$ with $\sigma=(\sqrt{5}-1)/2$ the
golden mean and, e.g., $d=29$ in Fig.~\ref{fig:Husimis}. This ensures
that there are no undesired periodicities and that $\heff$ is
approximately constant when varying $M$. Moreover, the
operator $\hat{U}$ reduces to an $N\times N$ periodic band
matrix. Using the symmetrized version of the map and making a unitary
transformation to a band matrix we are able to calculate the
eigenstates $|\psi\rangle$ of $\hat{U}$ up to $N \approx 
10^5$.

For $M=1$, Fig.~\ref{fig:Husimis}(c) shows a typical chaotic
eigenstate and five regular states.  The chaotic state extends over
the chaotic phase-space component and the regular states concentrate
on quantized tori. The eigenstates are represented on the classical
phase space by the Husimi distribution, where for visualization we use
tilted coherent states adapted to the shape of the island. For larger
system sizes we find that chaotic states flood the islands of
classically regular motion. Fig.~\ref{fig:Husimis}(a) shows such a
state for $M=1597$ that clearly ignores the outer tori of the island,
which for $M=1$ act as barriers for chaotic states
(Fig.~\ref{fig:Husimis}(c), left). In the central part of the island,
however, this state has essentially no weight. This partial flooding
of the island is observed even better in Fig.~\ref{fig:Husimis}(b),
where an average of the Husimi function of this state is taken over
all $M$ unit cells. The almost constant value in the chaotic component
extends well into the island.  Inside the island, clearly away from
its outer boundary, the Husimi function sharply drops to zero.

\begin{figure}[bth]
  \begin{center}
     \PSImagx{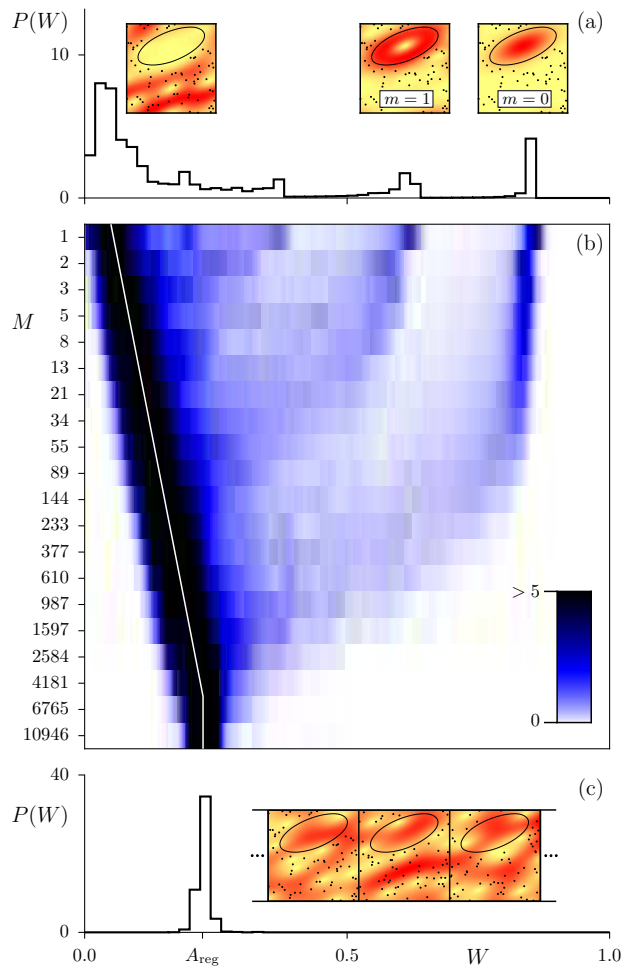}{8.2cm}
     \caption{Distribution $P(W)$ of the weights $W$ of the
     eigenstates in the regular islands for $\heff \approx 1/10$. (a)
     $M=1$: The main peak near $W=0$ is due to chaotic states and two
     further peaks are due to regular states $m=0, 1$ (see
     insets). (b) For increasing $M$ the main peak shifts to larger
     values of $W$ (white line) and the two peaks from the regular
     states disappear sequentially.  (c) $M=10946$: A narrow peak
     remains around $W = \Areg$. Three phase space cells of an
     eigenstate show a complete flooding of the islands.
     \label{fig:weight-distribution}} 
  \end{center}
\end{figure}

For a quantitative description of this flooding we now analyze the
weight $W$ of each eigenstate inside the islands. We determine this
weight by integrating the normalized Husimi function (calculated on a
$30 \times 30$ grid per unit cell) over the islands. In the
semiclassical limit, $\heff\to 0$, regular states have $W=1$, while
chaotic states have $W=0$. The distribution $P(W)$ of these weights
for all eigenstates is shown in Fig.~\ref{fig:weight-distribution} for
various system sizes and $\heff \approx 1/10$ \cite{footnote1}. For
$M=1$ we observe, as expected, a main peak near $W=0$ coming from the
chaotic eigenstates and two distinct peaks at larger $W$ from the two
regular states, see Fig.~\ref{fig:weight-distribution}(a). A
remarkable shift of the main peak of $P(W)$ to larger values of $W$
can be observed in Fig.~\ref{fig:weight-distribution}(b). This shows
that by enlarging the system size $M$ all chaotic states continuously
increase their weight inside the regular islands. This increase stops
when the center of the main peak reaches $W = 0.22$, which corresponds
to the area $\Areg$ of the island.  For these system sizes all states
completely flood the island (Fig.~\ref{fig:weight-distribution}(c),
inset), as observed in Ref.~\cite{HufKetOttSch2002}.

\begin{figure}[t]
  \begin{center}
     \PSImagx{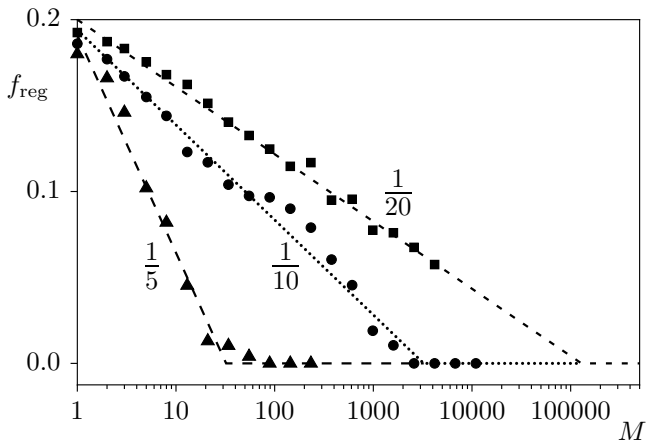}{8.6cm}
     \caption{Fraction $\freg$ of regular states vs.\ system size $M$
     for $\heff\approx 1/5$ (triangles), $1/10$ (circles), and $1/20$
     (squares). An approximately linear decrease with $\ln M$ to
     $\freg=0$ can be seen (lines are a guide to the eye). Already for
     small system sizes $M \approx 10$ and small $\heff$ a significant
     reduction of regular states is observed.
     \label{fig:fractions} } \end{center}
\end{figure}

What happens to the regular states as $M$ is increased? 
Figure~\ref{fig:weight-distribution}(b) shows that the corresponding
peaks in the distribution $P(W)$ disappear. Notably, the peak for
$m=0$ is much longer visible than the peak for $m=1$. The nemesis of
the regular states can be quantified by determining their fraction
$\freg$ as a function of $M$. To this end we define a state to be
regular when its weight $W$ inside the islands exceeds 50\%, where the
exact criterion does not affect our analysis.
Figure~\ref{fig:fractions} shows that the fraction $\freg$ decreases
from approximately $\Areg$ all the way to zero. The decay is slower
for smaller $\heff$.

Remarkably, in Fig.~\ref{fig:fractions} we see 
strong signatures of the decrease of $\freg$ already for small system
sizes $M \approx 10$.  This holds even for small values of $\heff$,
where the complete flooding of the island is numerically not
accessible.  Similarly, a shift of the main peak in
Fig.~\ref{fig:weight-distribution}(b) is clearly detectable for such
small systems. We thus find that partial flooding of regular islands
is easily observable.

Why do chaotic states flood the islands of regular motion and why do
regular states disappear as the system size is increased? Let us
consider a single regular island coupled by tunneling to a chaotic
sea.  If the chaotic sea is infinite, its states form a continuum. A
regular state on the $m$-th quantized torus has a decay rate
$\gamma_m$ to the continuum \cite{footnotedecay}. Thus, it is not an
eigenstate, but it is dissolved into chaotic states. As a consequence,
the chaotic states occupy the $m$-th quantized torus of the island. If
the chaotic sea is finite, but large enough, this decay of the $m$-th
regular state may still take place. The condition for the decay is
that during the time $1/\gamma_m$ the discrete chaotic spectrum is not
resolved, leading to $1/\gamma_m \ll h/\Dch = \tH$ \cite{FGR}.  On the
other hand, if the chaotic sea is so small that during the time
$1/\gamma_m$ the chaotic spectrum is well resolved, then the regular
state $m$ does not decay, yielding Eq.~(\ref{newcondition}).  Note,
that $\gamma_m$ increases monotonically with $m$, as for larger $m$
the $m$-th torus is closer to the boundary of the island.

The quantized tori of an island can thus be grouped into two classes:
(i) The innner tori, $m=0,...,\mstar-1$, where condition
(\ref{newcondition}) is fulfilled and regular states exist. (ii) The
outer tori, $m=\mstar, ...,\mmax-1$, where Eq.~(\ref{newcondition}) is
violated and which is flooded by chaotic states.  Here $\mmax$ is the
number of quantized tori at a given $\heff$. We find for the fraction
$\freg$ of regular states and the weight $\Wch$ of chaotic states
inside the island:
\begin{equation} \label{eq:freg-Wch}
  \freg=\Areg \frac{\mstar}{\mmax} , \quad 
  \Wch= \Areg \left( 1- \frac{\mstar}{\mmax} \right) .
\end{equation}

Variation of the system size $M$ in our example allows to change
the Heisenberg time $\tH \sim M$, 
while keeping the rates $\gamma_m$ fixed.
Enlarging $M$ leads via Eq.~(\ref{newcondition})
to a decrease of $\mstar$, 
starting from $\mstar=\mmax-1$ all the way to $\mstar=0$.
Together with Eq.~(\ref{eq:freg-Wch})
this explains Fig.~\ref{fig:weight-distribution}(b),
where the regular state with $m=1$ disappears
before the $m=0$ state 
and the weight $\Wch$ grows until $\Wch=\Areg$,
where the island is completely flooded, 
see Fig.~\ref{fig:weight-distribution}(c).
This also explains the decrease of $\freg$ from $\Areg$ to $0$,
as observed in  Fig.~\ref{fig:fractions}.
This decrease occurs over an exponentially large range
in $M$, due to  the roughly
exponential dependence of $\gamma_m$ on $m$.
A quantitative understanding requires a theory
for the decay rates $\gamma_m$,
which is a subject of current research on dynamical tunneling 
\cite{DynTun,PodNar2003}.
Note, that in the case of chaos-assisted tunneling
the splitting of symmetry related regular
states fluctuates strongly, depending on individual chaotic states.
In contrast, the decay rate $\gamma_m$ describes an average
tunneling to a continuum of chaotic states.

Variation of $\heff$ affects both $\gamma_m$ and $\tH$ in 
Eq.~(\ref{newcondition}). While $\tH\sim M/\heff$,
one expects in analogy to WKB theory that 
$\gamma_m \sim \exp[ - g(m/\mmax)/\heff]$,
where the system specific function $g$ decreases monotonically to $g(1)=0$.
From  the definition of $\mstar$ follows
$\mstar/\mmax = g^{-1}[\heff \ln(M/\heff)]$,
where $g^{-1}$ decreases monotonically.
Decreasing $\heff$ reduces the argument of $g^{-1}$
such that $\mstar/\mmax$ increases.
Eq.~(\ref{eq:freg-Wch}) implies that 
$\freg$ grows and $\Wch$ decreases.
Note that in the semiclassical limit, $\heff\to 0$,
we obtain $\mstar/\mmax\to 1$.
This is in agreement with the semiclassical eigenfunction
hypothesis, namely $\freg=\Areg$
and there is no flooding.
In contrast, if the system size is infinite, we have an infinite
$\tH$ and our argument leads to $\mstar=0$, i.e.\ complete flooding,
for any $\heff$. This coincides with the considerations of 
\cite{HufKetOttSch2002}
implying a failure of the semiclassical eigenfunction hypothesis.

Our explanation is complete for systems without localization. For
example, this is the case if the average classical drift of a unit
cell is non-zero, like in atom optic experiments in the presence of
gravity ~\cite{expAccel,theoryAccel}. Localization, however, sets a
lower bound to the effective mean level spacing, $\Dch \sim
1/\lambda$, where $\lambda$ is the localization length.  For $M >
\lambda$, this leads to $\tH \sim \lambda$ and $\mstar$ stays at its
value for $M=\lambda$.  According to Eq.~(\ref{eq:freg-Wch}) the same
holds for $\freg$ and $\Wch$. This applies, e.g., to dynamical
localization in the kicked rotor. For transporting islands, like in
the model studied here, $\lambda \sim 1/\gamma_0$ is unusually large
\cite{HanOttAnt1984,IomFisZas2002andrefs,HufKetOttSch2002},
such that already for $M = \lambda$ one has $\mstar=0$, $\freg=0$, and
$\Wch=\Areg$.  In this case, localization has no consequences
\cite{footnote2}.

\begin{figure}[tbp]
  \begin{center}
     \PSImagx{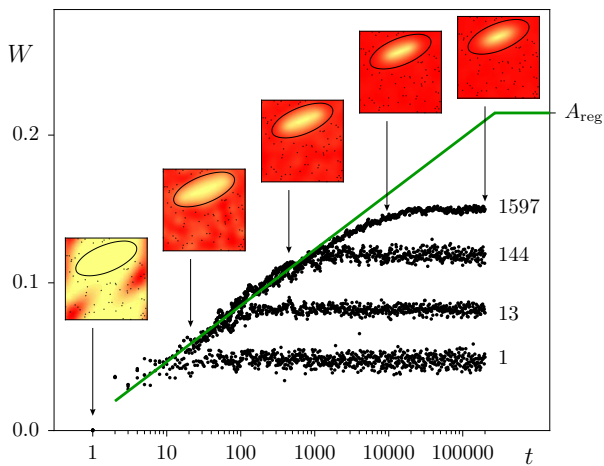}{8.0cm}
     \caption{Weight $W$ in the island vs.\ time~$t$ for a wave packet
     initially started as a coherent state in the chaotic sea at
     $(x,p)=(0.4,-0.2)$ for system sizes $M=1,13,144,1597$ and
     $\heff\approx 1/30$.  The thick (green) line guides the eye to
     the linear increase with $\ln t$ until $W=\Areg$. Insets show the
     time-evolved wave packet averaged over all cells for the case
     $M=1597$, demonstrating the progressive flooding of the island.
     \label{fig:temp-flooding} }
     \end{center}
\end{figure}

We generalize our analysis to the dynamics of wave packets, which is
experimentally of great relevance~\cite{expAccel}. A wave packet
started on the $m$-th torus will be restricted to that region if
condition (\ref{newcondition}) is fulfilled, i.e.\ $m<\mstar$.  If
$m>\mstar$, however, the wave packet will decay into the chaotic sea.
Particularly interesting is the case of a wave packet started inside
the chaotic sea. The island is progressively flooded, i.e.\ the $m$-th
torus at time $t_m \sim 1/\gamma_m$ for $m>\mstar$. For $t>\tH$ the
weight $W(t)$ will saturate at $\Wch$, Eq.~(\ref{eq:freg-Wch}).  This
is confirmed in Fig.~\ref{fig:temp-flooding}, for increasing values of
$M$.

Our results have consequences for spectral statistics in mixed systems
which go well beyond the previously studied effects of dynamical
tunneling (see e.g.\ \cite{ProRob94b,PodNar2003}).  The effective size
of the regular region, $\freg$ in Eq.~(\ref{eq:freg-Wch}), entering
the Berry-Robnik formula \cite{BerRob84} is drastically reduced.

Our analysis applies as well to hierarchical
states~\cite{KetHufSteWei2000}, which are confined by partial
transport barriers with turnstile areas smaller than $h$.  We predict
the additional condition $\gamma < 1/\tH$ for their existence, where
$\gamma$ describes the decay through these partial barriers. For
regular states on island chains within that hierarchical region,
condition (\ref{newcondition}) applies, with $\tH$ given by the mean
level spacing of the surrounding hierarchical states.

Finally, we emphasize that the time periodicity of the system
(\ref{hamiltonian}) and the restriction of our discussion to maps is
not crucial and that we expect flooding of islands for any Hamiltonian
with a mixed phase space. We stress that this new quantum signature of
chaos for eigenstates and wave packet dynamics already appears for
small system sizes, e.g.\ island chains of length 10. This makes
numerical explorations very feasible and should lead to experimental
observations, for example using optical
lattices~\cite{SteOskRai2001,Hen2001,expAccel}.

We thank Holger Schanz for discussions and the DFG for support under
contract KE 537/3-2.


\begin{thebibliography}{10}
\bibitem{footnote1}
We improve the statistics for small $M$ by varying the phase of the
boundary condition in the $x$-direction, which does not affect the
underlying classical dynamics. For large $M$ we use 2000 eigenstates.
\end{thebibliography}
\end{document}